\documentclass[aps,showpacs,prd,toolkits,twocolumn, superscriptaddress]{revtex4}

\usepackage{amsmath}    
\usepackage{graphicx}   
\usepackage{verbatim}   
\usepackage{color}      
\usepackage{subfigure}  
\usepackage{hyperref}   

\newcommand{\beq}{\begin{equation}}   
\newcommand{\eeq}{\end{equation}}
\newcommand{\beqn}{\begin{eqnarray}}   
\newcommand{\eeqn}{\end{eqnarray}}

\newcommand{\ra}{\rightarrow}

\newcommand{\gsim}{\lower.7ex\hbox{$
\;\stackrel{\textstyle>}{\sim}\;$}}
\newcommand{\lsim}{\lower.7ex\hbox{$
\;\stackrel{\textstyle<}{\sim}\;$}}

\def\Xint#1{\mathchoice
   {\XXint\displaystyle\textstyle{#1}}%
   {\XXint\textstyle\scriptstyle{#1}}%
   {\XXint\scriptstyle\scriptscriptstyle{#1}}%
   {\XXint\scriptscriptstyle\scriptscriptstyle{#1}}%
   \!\int}
\def\XXint#1#2#3{{\setbox0=\hbox{$#1{#2#3}{\int}$}
     \vcenter{\hbox{$#2#3$}}\kern-.5\wd0}}

\def\dashint{\Xint-}

\begin{document}
\title{How Chiral Symmetry Breaking Affects the Spectrum of the Light-Heavy Mesons\\ in the 't Hooft Model}
\author{L.Ya.~Glozman}
\affiliation{Institut f\"{u}r Physik, FB Theoretische Physik, Universit\"{a}t Graz, Universit\"{a}tsplatz 5, A-8010 Graz, Austria}
\author{V.K.~Sazonov}
\altaffiliation{Also at the Department of Theoretical Physics, St. Petersburg University, 
Uljanovskaja 1, St. Petersburg, Petrodvorez, 198504 Russia}
\affiliation{Institut f\"{u}r Physik, FB Theoretische Physik, Universit\"{a}t Graz, Universit\"{a}tsplatz 5, A-8010 Graz, Austria}
\author{M.~Shifman}
\affiliation{William I. Fine Theoretical Physics Institute, University of Minnesota, Minneapolis, MN 55455, USA}
\author{R.F. Wagenbrunn}
\affiliation{Institut f\"{u}r Physik, FB Theoretische Physik, Universit\"{a}t Graz, Universit\"{a}tsplatz 5, A-8010 Graz, Austria}

\begin{abstract}
We find the spectrum and wave functions of 
the heavy-light mesons in  $(1 + 1)$-dimensional QCD in the 't Hooft limit, both  in the rest frame,
using the Coulomb (axial) gauge, and on the light cone. Our emphasis is on the
 effects of chiral symmetry breaking on the spectrum.
While dynamical equations in both cases look different, the results for the   spectrum
are identical. The chiral symmetry breaking is clearly seen from the gap and Bethe--Salpeter equations
in the laboratory frame. At the same time, while vacuum is trivial on the light cone (no chiral condensate), 
the effects of the spontaneous 
breaking of the chiral symmetry manifest themselves in  the same way, as it follows from the coincidence  of the  spectra obtained from the laboratory-frame Bethe--Salpeter equation on the one hand, and the light-cone 't Hooft-type equation on the other.
\end{abstract}


\maketitle

\section{Introduction}
\label{intro}

In this paper we discuss the relation between the chiral symmetry breaking 
in the two-dimensional 't Hooft model \cite{tHoof1} 
and the heavy-light meson mass spectrum. 

The action of the version of the 't Hooft model we will consider
 is
\beq
S  = \int d^2 x\left[ -\frac{1}{4} G_{\mu\nu}^aG_{\mu\nu}^a
+ 
\sum_{f=1,2} \bar \psi_f ( iD \hspace{-0.37cm}\not \,\,\,\,  -m_f )\psi_f\right]
\eeq
where 
$G_{\mu\nu}^a$ is the gluon field strength tensor, the index $a$ runs 
from $1$ to $N^2 -1 $, and $N$ is the number of colors, 
$$
N\rightarrow\infty\, .
$$
The subscript $f$ marks quarks of different flavors. The quarks 
are assumed to belong to the fundamental representation of the 
gauge group SU$(N)$. 
Moreover, in our consideration we will assume that $m_2\to\infty$,
so that the second quark will play the role of a static force center, while
$m_1\to 0$ so that the first quark is massless. The theory then possesses two U(1) symmetries, 
generated by the vector and axial currents, $\bar\psi_1\gamma^\mu\psi_1$ and 
$\bar\psi_1\gamma^\mu\gamma^5\psi_1$, respectively. 
The axial symmetry is spontaneously broken (see below).

The coupling constant $g$ has dimension of mass,  and 
in the large-$N$ limit scales as
\beq
\lambda \equiv \frac{g^2 N}{4\pi} = {\rm const}\, .
\eeq
The constant $\lambda$ is referred to as the 't Hooft coupling.

The very fact of confinement is obvious in this model since in two 
dimensions the Coulomb potential generated by the static color source
(i.e. the infinitely heavy quark at the origin) grows linearly with
separation.  The model was  solved in the 
light-cone formalism by 't Hooft \cite{tHoof1}  and further
developed along the same lines in Refs. \cite{Call1,Einh1}.
The spectrum of the light-light mesons  and the light-cone 
 wave functions were obtained from the 't Hooft equation,
 an integral equation, supplemented by certain boundary 
conditions,  well studied in the literature (for a review see e.g. \cite{Hornbostel:1988ne}).

In the light-cone formalism one chooses the light-cone gauge 
condition,
$$
A_- = 0 \, .
$$
The light-cone time derivative of $A_+$ does not appear in
$G_{\mu\nu}$; hence, $A_+$ is a non-dynamical degree of freedom 
which can be eliminated through the equations of 
motion. In the large-$N$ limit the only surviving diagrams
are ladders and rainbows. The 't Hooft equation
for the  bound state built from the quark of the first flavor and 
anti-quark of the second flavor has the form
\beq
\left( \frac{m_1^2}{x} + \frac{m_2^2}{1-x} - M^2\right) \phi (x)
=2 \lambda \,\,\int_0^1\, 
\frac{\phi (y) - \phi (x) }{(x-
y)^2}dy
\, ,
\label{thooftequ}
\eeq
where $x$ is the first quark's share of the total (light-cone) momentum
of the composite meson with mass $M$. 
If we deal with  massless (anti)quarks in the equation above ($m_1=m_2=0$),
Eq. (\ref{thooftequ})  has a massless-meson solution (``pion" with $M=0$)
which  is known 
exactly. The corresponding  light-cone wave function is $x$-independent,
 $ \phi (x) = $const.  The existence of the massless pion implies \cite{Zhit1}, 
through the standard current algebra relations, a non-vanishing 
quark condensate \footnote{ 
Generally speaking, in two dimensions any continuous (e.g. chiral)
 symmetry cannot be spontaneously broken (which is known as the
 Mermin-Wagner-Coleman theorem).
This is because  massless Goldstone bosons would bring a long-range
infrared divergence for $d=2$. However,  
at $N_c=\infty$  self-interaction of Goldstone bosons 
vanishes (they do not interact also with all other mesons) 
and, consequently, at $N_c=\infty$ the chiral symmetry can and is
indeed spontaneously broken in the 't Hooft model.}
$
\langle  \bar \psi \psi \rangle$ proportional to  $
- N \sqrt \lambda$, see also 
\cite{Ming1,Lenz1,Burk1}.
The problem 
is that this chiral condensate is not seen directly in the light-cone 
consideration, a usual story with all light-cone analyses 
of the vacuum condensates.
The chiral condensate on the light cone 
 is buried somewhere in zero modes and boundary conditions.
 
Indeed, if one tries to extract the quark condensate
directly from the light-cone quark Green's function given by 't~Hooft,
one obtains
\beq
\langle \bar \psi \psi \rangle \propto \lim_{x\ra 0}\, {\rm Tr}
\left\{ S(x,0)\right\}\, ,
\label{GF}
\eeq
where $S(x,0)$ is the massless quark Green's function
describing the quark propagation from the point 0 to 
the point $x$. The 
 right-hand side vanishes after taking  trace, since this
Green's function is linear in the $\gamma$ matrices. 

Our task is not only to reveal the chiral condensate (this had been already done 
by shifting slightly away from the light cone \cite{Lenz1} or,  from the solution
of the gap equation in the laboratory frame \cite{Bars1}), but also to analyze its impact on the spectrum
of  bound mesons. In order to keep a closed-form integral equation 
\'a la 't Hooft as the spectral equation we
have to focus on a system of an infinitely heavy anti-quark at rest
at the origin and a dynamical quark of mass $m_1\to 0$ bound by a linearly growing 
potential, i.e. the heavy-light quark system. The bound quark is ultra-relativistic,
and dynamical details of its binding crucially depend  on the
chiral condensate (see below). 
At the same time, the system in question can be considered in the laboratory frame 
(as opposed to the light-cone consideration). The static infinitely heavy (anti)quark suppresses
the so called $Z$ graphs in much the same way as the
transition to the light cone in the case of two massless (anti)quarks. 
The absence of the $Z$ graphs is necessary to
keep the spectral equation in the closed form. 
The above integral equation applies to
the one-particle wave function in the momentum space. 
It can be readily obtained from the general analysis
of \cite{Bars1} in the limit $m_2\to \infty$ and $m_1\to 0$. We will briefly review the derivation below.

Another aspect, to be addressed below, is the 
the relation with the ``original" light-cone
spectral equation for the heavy-light system, which we will refer to as the
't Hooft-like equation. It was obtained \cite{Bur,Zhit2} from the general light-cone 't~Hooft equation valid for arbitrary
$m_{1,2}$ in the limit $m_2\to\infty$ and $m_1\to 0$.
In fact, we deal with {\em two different} one particle 
equations. One of them is just a limiting case of the 't Hooft equation, 
and applies to the light-cone wave function,
which depends on $x$ ($0\leq x\leq 1$). Within this approach the (massless quark) 
condensate vanishes. 
At the same time, our laboratory frame equation has the condensate buit in.
It is the spectral equation for $\phi(p)$ where $p$ is the light-quark momentum in the laboratory frame.
In deriving these two equations one uses two distinct limiting procedures. 
To obtain the 't Hooft-like equation one first tends the momentum to infinity, keeping the quark masses fixed, and then tends one of the quark masses to infinity.
At the same time, when one works in the laboratory frame, one keeps the total momentum fixed and sends the
quark mass to infinity from the very beginning. Generally speaking, these two limits need not be commutative.

Our analysis will demonstrate that the above two equations are, in fact, {\em isospectral};
i.e. the limiting procedures are interchangeable, with no obstructions.

Surprisingly, the
laboratory frame equation for $\phi (p)$ {\em formally} becomes identical to
the 't' Hooft-like equation for 
$\varphi (\xi )$ (see Eq. (\ref{HQLtHooft})) upon substitution into the laboratory-frame equation
a ``wrong" solution for the chiral angle (i.e. a singular solution with no chiral symmetry breaking) and a rescaling of the overall
energy scale. This curious coincidence has no obvious physical reason; at least, we were unable to
find such a reason.

The heavy-light systems in the 't Hooft model were considered
previously, in an applied context, e.g. in Ref. \cite{Grin1}. In this work 
the original light-cone 't Hooft equation was numerically solved at large values of
$m_2/\sqrt\lambda$. As was mentioned, in the 't Hooft-like equation the limit 
$m_2/\sqrt\lambda \longrightarrow \infty$ is taken {\em before} solving the 't~Hooft equation.
The appropriate limiting procedure was implemented in  \cite{Bur,Zhit2}.
Note that when the heavy-light meson is boosted (to put it on the light cone)
the total momentum of the meson is shared between quarks
proportionally to their masses. Therefore, the heavy quark will have 
$x$ very close to unity while the light quark's share will be close to 
zero. The width of the $x$ distribution will be proportional to
$\sqrt\lambda/m_2\to 0$. This fact was noted long ago
\cite{Bjor1}, and was later extensively exploited in  phenomnenology. 
The light-cone wave function will have an infinitely narrow support
  in the limit $\sqrt\lambda/m_2\ra 0$ unless we rescale the variable $x$,
so that 
the corresponding distribution does not shrink to a delta function but 
is, rather, characterized by a constant width. 

The appropriate rescaling laws are as follows \cite{Bur,Zhit2}:
\begin{eqnarray}
x &=&1- \frac{\sqrt{2 \lambda}}{m_2}\xi\, ,\nonumber\\[2mm]
M &=& m_2 + {\mathcal E}\, ,
\nonumber\\[2mm]
\phi (x ) &=& \sqrt{m_2 (2 \lambda )^{-1/2}}\,\,  \varphi (\xi )\, ,
\end{eqnarray}
 where $m_2$ is to be sent to infinity while ${\mathcal E}$ is
kept fixed (i.e. ${\mathcal E}$ is the mass of the bound state after the subtraction of the
mechanical mass of the infinitely heavy anti-quark).


Then the light-cone 't Hooft equation takes the form  
\begin{equation}
  2 {\cal E} \varphi(\xi) = \sqrt{2 \lambda}\, \xi\, \varphi(\xi) - \sqrt{2 \lambda}
  \int_0^\infty \, 
  \frac{\varphi(\tilde{\xi}) - \varphi(\xi)}{(\tilde{\xi} -
  \xi)^2}d\tilde{\xi}\, .
\label{HQLtHooft}
\end{equation}
The boundary conditions in this equation are as follows:
\begin{equation}
  \varphi(\xi \rightarrow 0) \rightarrow {\rm const}\, ,~~\varphi(\xi \rightarrow \infty) \rightarrow 0\, .	\label{boundcondZ}
\end{equation}


Our main results can be summarized as follows.
We solve the heavy-light system in the laboratory frame using
the Coulomb (axial) gauge. As the first step we solve the
gap equation and obtain the required quark Green function.
Given this quark Green function we are in position to solve
the Bethe--Salpeter equation. Both the single-quark Green function
(the quark condensate follows straightforwardly from the quark
Green function) and the meson spectrum manifestly
exhibit dynamical chiral symmetry breaking. Then we solve the
same system on the light cone by integrating (numerically)  the 't Hooft-like equation.
We obtain exactly the same spectrum even though the dynamical equations
in both cases have very different physical meaning, and there is no gap equation
on the light cone. Dynamical chiral symmetry breaking is manifest through
the absence of parity doubling in the spectrum in both cases, but
in the laboratory frame this chiral symmetry breaking is also clearly
seen through the nonzero quark condensate in the vacuum. While
all the intermediate color-nonsinglet quantities, such as the
quark Green function, manifestly depend on the reference frame and on
the gauge-fixing condition, the spectrum of the color-singlet
system is independent of the choice of the quantization scheme,
of the reference frame and of the gauge condition.

In Section \ref{ChiralSym} we briefly review the chiral symmetry breaking and solution of the associated gap equation in the
laboratory frame. In Section \ref{hsmes} we discuss the spectral equation for the heavy-light mesons in the
laboratory frame and on the light cone. Numerical solutions are presented. Section \ref{concl} briefly summarizes our results and conclusions.

\section{\label{ChiralSym} Chiral symmetry breaking in  vacuum} 

\subsection{The gap equation}

In the laboratory frame,  the axial (Coulomb)  gauge condition 
\begin{equation}
A_1 = 0
\label{gaugefix}
\end{equation}
is convenient.
The derivation of the bound state equation is  carried 
out in two  steps, see  \cite{Bars1} for details.  First one  
needs to obtain  the
 quark Green's function for the massless quark. Its self-energy 
 saturated in the large-$N$ limit by the 
rainbow graphs.

\begin{figure}[!htbp]
  \begin{center}
    \includegraphics[width=0.3\linewidth]{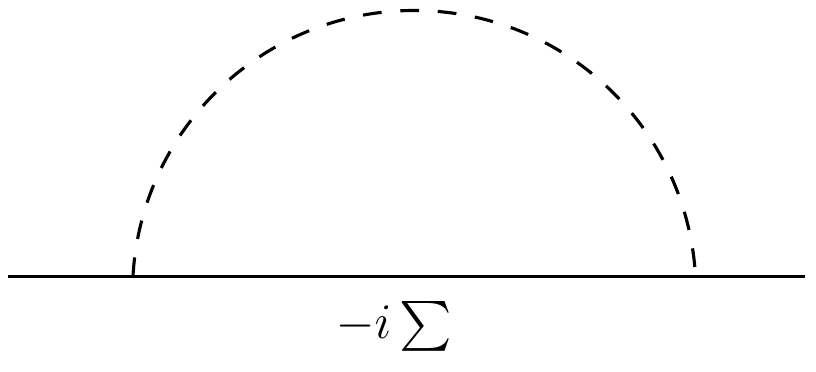}
\caption{\small 
Quark self-energy at one loop.}
\label{tf13}
  \end{center}
\end{figure}

To introduce necessary notation  it is convenient to start,
however,  from  the one-loop graph
presented in Fig.~\ref{tf13}.


We will denote the quark self-energy by $-i \Sigma $, so that
the quark Green's function is
\beqn
G_{ij}(p_0,p) &=&\int d^2 x\, e^{ip_\mu x^\mu}\,\left\langle T\left\{
\psi (x)\, \bar\psi (0)\right\}\right\rangle \nonumber\\[3mm]
&=& \frac{i}{\not\! p -m -\Sigma}\,,
\label{qgfo}
\eeqn
where the mass parameter $m$ is arbitrary (real and positive) for the time being.
In the $A_1=0$ gauge  
$\Sigma$  depends only on the
spatial component of the quark momentum $p$, not on $p^0$.
  In  calculating the graph of Fig.
\ref{tf13}  we benefit from the fact that only $D_{00}$ is non-vanishing,
and perform the integral over the time component of the 
loop momenta using residues. In this way we arrive at
\begin{widetext}
\beqn
\Sigma (p) &=&\frac{\lambda}{2}\left\{
-2\gamma^1\left[\frac{p}{m^2+p^2}+\frac{m^2}{2(m^2+p^2)^{3/2}}
\ln\frac{\sqrt{m^2+p^2} +p}{\sqrt{m^2+p^2} -p}\right]\right.\nonumber\\[4mm]
&-&
m\left.\left[\frac{2}{m^2+p^2}-\frac{p}{(m^2+p^2)^{3/2}}
\ln\frac{\sqrt{m^2+p^2} +p}{\sqrt{m^2+p^2} -p}\right]
\right\}.
\label{qgfu}
\eeqn
\end{widetext}
Now we see that
(i) The loop expansion parameter is $\lambda/(m^2+p^2)$;
it explodes at $m,p <\sqrt\lambda$, so that summation 
of the infinite series is necessary;
(ii) In the $A_1=0$ gauge $\Sigma$ depends only on the spatial
component of momentum; (iii) Its general Lorentz structure is
\beq
\Sigma (p) = A(p) + B(p) \,\gamma^1\,,
\eeq 
where $A$ and $B$ are some real functions of $p$ (for real $p$). From Eq.~(\ref{qgfo})
we see that the combination we will be dealing with 
in the quark Green's  function is
\beq
m +p\, \gamma^1 +A(p) + B(p) \,\gamma^1\,.
\label{dvap}
\eeq
Usually $A$ and $B$ are traded for two other functions,
which  parametrize the quark Green's function in a more convenient way. Namely,
\beqn
&& E_p \equiv \sqrt{(m+A(p))^2+(p+B(p))^2}\,,\nonumber\\[3mm]
&& m+A(p) = E_p\,\cos\theta_p\,,\nonumber\\[3mm] && p+B(p)= E_p\,\sin\theta_p\,,
\label{dvapp}
\eeqn
where for consistency one should demand $E_p$ to be {\em positive} for all real $p$.
The angle $\theta_p$ is referred to as the Bogoliubov angle, or, more commonly,
the chiral angle. The exact quark Green's function now can be rewritten
as
\beq
G = i\,\,\frac{p^0\gamma^0 - E_p\,\sin\theta_p\,\gamma^1 + E_p\,\cos\theta_p}{p_0^2-E_p^2+i\varepsilon}\,.
\label{egfp}
\eeq

\begin{figure}[!htpb]
  \begin{center}
    \includegraphics[width=0.3\linewidth]{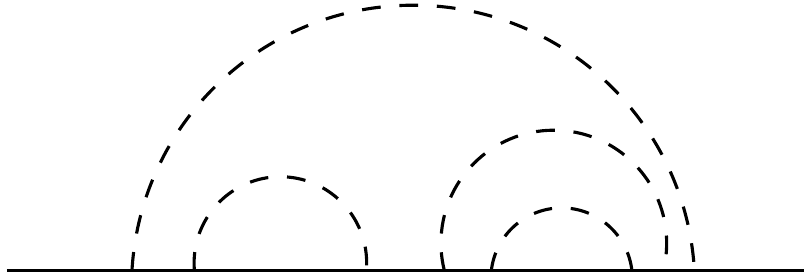}
\caption{\small 
An example of the rainbow graph in $\Sigma (p)$. }
\label{tf14}
  \end{center}
\end{figure}

\begin{figure}[!htpb]
  \begin{center}
    \includegraphics[width=0.3\linewidth]{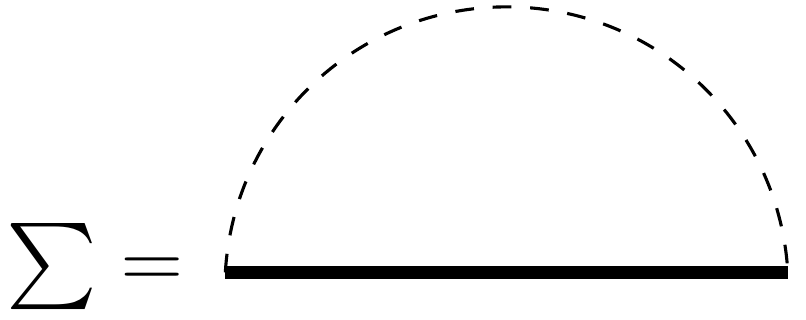}
\caption{\small 
Exact equation for $\Sigma (p)$ summing all rainbow graphs. The bold solid line
is the exact quark propagator (\ref{egfp}).}
\label{tf15}
  \end{center}
\end{figure}

Closed-form exact equations can be obtained for
$E_p$ and $\theta_p$ due to the fact that in the 't Hooft limit
the quark self-energy is saturated by ``rainbow graphs."
An example of the rainbow graph is depicted in Fig.~\ref{tf14}.
Intersections of the gluon lines and insertions of the internal quark loops are
forbidden, and so are the gluon lines on the other side of the quark line. This diagrammatic structure implies an equation  
depicted in Fig.~\ref{tf15}, where the bold solid line denotes the exact Green's function
(\ref{egfp}). 
Algebraically
\beq
\Sigma (p) = \frac{i\,\lambda}{2 \pi}\,\dashint\,\frac{d^2k}{(p-k)^2}\,
\gamma^0 G(k)\gamma^0\,.
\label{SigmaPV}
\eeq
It is easy to see that this equation sums up
the infinite sequence of the rainbow graphs in its entirety. 
In Eq. (\ref{SigmaPV}) a principal value of the integral on the right-hand side is assumed.

Using  
(\ref{egfp}) and performing integration over $k^0$, the time component
of the loop momentum, by virtue of residues, it is not difficult to obtain
\beq
\Sigma (p) =\frac{\lambda}{2}\,\dashint dk\left\{\gamma^1\,\sin\theta_k\,\frac{1}{(p-k)^2}+ \cos\theta_k\,\frac{1}{(p-k)^2}
\right\}\,,
\eeq
which implies, in turn,
\beqn
&& A(p) = E_p\cos\theta_p - m = \frac{\lambda}{2}\,\dashint dk\, \cos\theta_k\,\frac{1}{(p-k)^2}\,,\nonumber\\[3mm]
&& B(p) = E_p\sin\theta_p - p = \frac{\lambda}{2}\,\dashint dk\, \sin\theta_k\,\frac{1}{(p-k)^2}
\,.
\label{eqs30}
\eeqn
This should be supplemented by the boundary conditions
\beq
\theta_p \to\left\{\begin{array}{ll}
\frac{\pi}{2}\quad{\rm at} \quad p\to \infty ,\\[2mm]
-\frac{\pi}{2}\quad{\rm at} \quad p\to -\infty ,
\end{array}
\right.
\label{eqs31}
\eeq
determined by the free-quark limit.
The integrals (\ref{SigmaPV}) -- (\ref{eqs30}) contain singularity at $p = k$,
so a regularization is required. We use the principal value regularization.
This set of equations, called the gap or the Schwinger--Dyson equation, was first obtained by Bars and Green \cite{Bars1}.
Multiplying the first equation by $\sin\theta_p$ and the second by
$\cos\theta_p$ and subtracting one from another one gets
an integral equation for the chiral angle, namely,
\beq
p\,\cos\theta_p - m\,\sin\theta_p =  \frac{\lambda}{2}\,\int dk\, \sin
(\theta_p-\theta_k)\,\frac{1}{(p-k)^2}\,.
\label{chirang}
\eeq
The latter equation, in contrast to (\ref{SigmaPV}) -- (\ref{eqs30}),
does not contain singularity at $p = k$.
Assuming that the chiral angle is found in the limit $m=0$
from 
\beq
 p\,\cos\theta_p  =  \frac{\lambda}{2}\,\int dk\, \sin
(\theta_p-\theta_k)\,\frac{1}{(p-k)^2}\,,
\label{chirangm}
\eeq
one can get  $E_p$ from the equation
\beq
 E_p =  p\,\sin\theta_p +
\frac{\lambda}{2}\,\dashint dk\, \cos (\theta_p-\theta_k)\,\frac{1}{(p-k)^2}\,.
\label{chirangmp}
\eeq
An immediate consequence is that $\theta_p$ is an odd function of $p$,
while $E(p)$ is even.

By solving the gap equation one obtains the chiral angle $\theta_p$ and both
dressing functions $A(p)$ and $B(p)$. In the chiral limit $m=0$ the chiral
symmetry breaking part of the quark Green function is $A(p)$. Consequently
a nonzero $A(p)$ signals dynamical chiral symmetry breaking in the vacuum.
It is an intrinsically non-perturbative effect that cannot be obtained
within the perturbation theory.

\subsection{A wrong solution}

Upon examining Eq.~(\ref{chirangm}) it is not difficult to
guess an analytic solution,
\beq
\theta_p = \frac{\pi}{2}\, {\rm sign}\,p\,,
\label{unsst}
\eeq
where  ${\rm sign}\,p$ is the sign function,
$$
{\rm sign}\,p =\vartheta (p) - \vartheta (-p)\,.
$$
The solution (\ref{unsst}) is singular. If nevertheless we use it, then 
substituting (\ref{unsst}) in Eq.~(\ref{chirangmp})
one obtains
\beq
E_p = |p|-\frac{\lambda}{|p|}\,\,.
\label{unsse}
\eeq
The above results shows that the  analytic solution (\ref{unsst})  is unphysical. This is obvious from the
fact that $E_p$ becomes negative at $|p|<\sqrt\lambda$.
This feature of the solution (\ref{unsse}) --- negativity at small
$|p|$ --- cannot be amended by a change of the infrared regularization. See also \cite{NefKalash}.

The unphysical solution (\ref{unsse}) leads to the vanishing quark condensate, as will be clear from Eq.
(\ref{psibarpsii}). We will return to the unphysical solution  later, after discussing the (nonsingular) physical solution.
 

\subsection{Physical solution}

A solution that leads to a nonvanishing condensate
has the form depicted in Fig. \ref{changle}.
It is smooth everywhere.
At $|p|\ll\sqrt\lambda$ it is linear in $p$. Its asymptotic approach to
$\pm\pi/2$ at
 $|p|\gg\sqrt\lambda$ will be discussed later.

Now, let us calculate the chiral condensate,
the vacuum expectation value $\langle\bar\psi\psi\rangle$,
\beq
\langle\bar\psi\psi\rangle = -\,{\rm Tr}\, \int \frac{d^2p}{(2\pi)^2}\,G(p_0,p)\, ,
\label{psibarpsi}
\eeq
where  Tr stands for both traces, with respect to color and Lorentz indices, and the quark Green
function $G(p_0,p)$ is defined in Eq.~(\ref{egfp}). Taking the trace and performing the $p_0$ integration we arrive at
\beq
\langle\bar\psi\psi\rangle = - N\,\int \frac{dp}{2\pi}\,\cos\theta_p\,.
\label{psibarpsii}
\eeq
For the singular solution (\ref {unsst}) the above quark condensate vanishes since
$\cos\theta_p\equiv 0$. However, for the physical smooth solution depicted in 
Fig. \ref{changle}
the   quark condensate does not vanish,
\beq
\langle\bar\psi\psi\rangle = - \frac{N}{\sqrt 6}\,\sqrt\lambda\,.
\label{zhit}
\eeq
 
  Equation~(\ref{psibarpsii})  
 in conjunction with  (\ref{chirangm}),  
allow us to determine the leading preasymptotic correction in
$\theta_p$ at $|p|\gg \sqrt\lambda$. Indeed, in this limit
the right-hand side of 
Eq.~(\ref{chirangm}) reduces to (at $p>0$)
\beq
\frac{\lambda}{2 p^2}\, \int dk\, \sin\left(\frac{\pi}{2}-\theta_k\right)=
\frac{\lambda}{2 p^2}\, \int dk\, \cos\theta_k\,,
\eeq
while the left-hand side
\beq
p\, \sin\left(\frac{\pi}{2}-\theta_p\right)\to p\, \left(\frac{\pi}{2}-\theta_p\right).
\eeq
This implies, in turn, that
\beq
\theta_p =\frac{\pi}{2}\, {\rm sign}\, p - \frac{\pi}{\sqrt 6}\left(\frac{\sqrt\lambda}{p}
\right)^3+ ...\,,\qquad |p|\gg\sqrt\lambda\,.
\eeq
At the same time, from Eq.~(\ref{chirangmp}) we deduce that there is no $p^{-3}$ correction in
$E/|p|$, the leading correction is of order of $\lambda^3/p^6$.

\subsection{\label{NGapSolSec} Numerical solution of the gap equation and an alternative scheme of regularization}

The gauge choice (\ref{gaugefix}) for the model (1) ensures the existence of only 
one non-trivial component of the gluon propagator:
\begin{eqnarray}
  \nonumber
  D_{01}^{ab}(x_0 - y_0, x - y) = D_{11}^{ab} (x_0 - y_0, x - y) = 0\, , \\[2mm]
  D_{00}^{ab}(x_0 - y_0, x - y) = -\frac{i}{2}\delta^{ab} |x - y| \delta(x_0 - y_0)\, .
\label{Dprop}
\end{eqnarray}
$D_{00}^{ab}(x_0 - y_0, x - y)$ corresponds to an instantaneous linear confining potential.
All loop integrals calculated with a linear potential diverge in the infrared region, 
hence one has to introduce an infrared regularization. This can be done in a number of ways. 
In previous sections we used a principal value regularization.

Here we apply an alternative regularization,
which suppresses the small momenta of the linear potential
by introducing a cutoff parameter into the propagator in the momentum representation. 
We define propagator in momentum representation as
\begin{equation}
  D_{00}^{ab}(x_0 - y_0, p) = i \frac{\delta^{ab} \delta(x_0 - y_0)}{p^2 + \mu_{IR}^2}\, .
\label{scdef}
\end{equation}
Then in the final answer for the color-singlet quantities the infrared limit $\mu_{IR} \rightarrow 0$ must be taken.

In the regularization scheme defined by (\ref{scdef}) the expression for the self-energy operator
(\ref{SigmaPV}) turns into
\begin{eqnarray}
  \Sigma(p) &=& \frac{\lambda}{2} \int dk \left[ \gamma^{1} \sin\theta_k \frac{1}{(p - k)^2 + \mu_{IR}^2} 
\right. \nonumber\\
  &+& \left. \cos\theta_k \frac{1}{(p - k)^2 + \mu_{IR}^2} \right]\, .
\label{Sigmamu}
\end{eqnarray}
Using the representation of the delta-function 
\begin{equation}
  \delta(x) = \lim_{\mu_{IR} \rightarrow 0} \,\frac{1}{\pi} \,\frac{\mu_{IR}}{x^2 + \mu_{IR}^2}\, ,	\label{deltaf}
\end{equation}
it is easy to see that the self-energy defined in (\ref{Sigmamu}) diverges at $\mu_{IR} \rightarrow 0$ as:
\begin{equation}
  \lim_{\mu_{IR} \rightarrow 0} \Sigma(p) = \frac{\lambda \pi}{2 \mu_{IR}} \sin\theta_p \gamma^1 + 
  \frac{\lambda \pi}{2 \mu_{IR}} \cos\theta_p + {\rm a~finite~part}\,.
\label{Sigmalim}
\end{equation}
The self-energy operator defined in (\ref{SigmaPV}) via the principal value regularization is always finite. 
This is also true for the energy of a single quark which, being regularized through
 (\ref{scdef}) takes the form
\beq
 E_p =  p\,\sin\theta_p +
\frac{\lambda}{2}\,\int dk\, \cos (\theta_p-\theta_k)\,\frac{1}{(p-k)^2 + \mu_{IR}^2} \, .
\label{Epmu}
\eeq
$ E_p$ diverges at $\mu_{IR} \rightarrow 0$ as
\beq
  \lim_{\mu_{IR} \rightarrow 0} E_p =  \frac{\lambda \pi}{2 \mu_{IR}} + {\rm finite~terms}\, ,	\label{Epbeh}
\eeq
while with the principal value regularization it is always finite.
For any other color-nonsinglet quantity one has the same situation.

This circumstance reflects the confining properties of the 't Hooft model.
Confinement means that only observable color-singlet quantities have finite well-defined
values, that should not depend on the infrared regularization scheme. The color-nonsinglet quantities
are not observable and manifestly depend on the regularization choice. Our present regularization
is convenient in the sense that it explicitly removes all color-nonsiglet objects from the physical Hilbert
space since they are all infrared divergent. At the same time this infrared divergence exactly cancels in all color-singlet
observable quantities, such as the meson spectrum, the chiral angle and the quark condensate.
The color-singlet quantities are finite and do not depend on the choice of the regularization.

In the following we show that the infrared divergences exactly cancel in the gap equation,
written in the form
\begin{equation}
  A(p)\,\sin\theta_p - [B(p) + p]\,\cos\theta_p = 0\, ,
\label{ABgapeq}
\end{equation}
where $A(p)$ and $B(p)$ in the regularization scheme (\ref{scdef}) are
\begin{eqnarray}
\nonumber
  A(p) = \frac{\lambda}{2} \int dk \frac{\cos\theta_k}{(p - k)^2 + \mu_{IR}^2}\, ,\\
  B(p) = \frac{\lambda}{2} \int dk \frac{\sin\theta_k}{(p - k)^2 + \mu_{IR}^2}\, .
\label{ABdefmu}
\end{eqnarray}
Using the representation of the delta function (\ref{deltaf}) we obtain at $\mu_{IR} \rightarrow 0$:
\begin{eqnarray}
\nonumber
  A(p) = \frac{\lambda \pi}{2 \mu_{IR}} \cos\theta_p + A_{\rm finite}(p)\, ,\\
  B(p) = \frac{\lambda \pi}{2 \mu_{IR}} \sin\theta_p + B_{\rm finite}(p)\, .
\label{ABfindivp}
\end{eqnarray}
Note that in (\ref{ABgapeq}) all divergences exactly cancel 
and
\begin{equation}
  \tan\theta_p = \frac{B(p) + p}{A(p)} = \frac{B_{\rm finite}(p) + p}{A_{\rm finite}(p)}\,.
\label{tanth}
\end{equation}
%
%

\begin{figure}[!htpb]
  \begin{center}
    \includegraphics[width=0.8\linewidth]{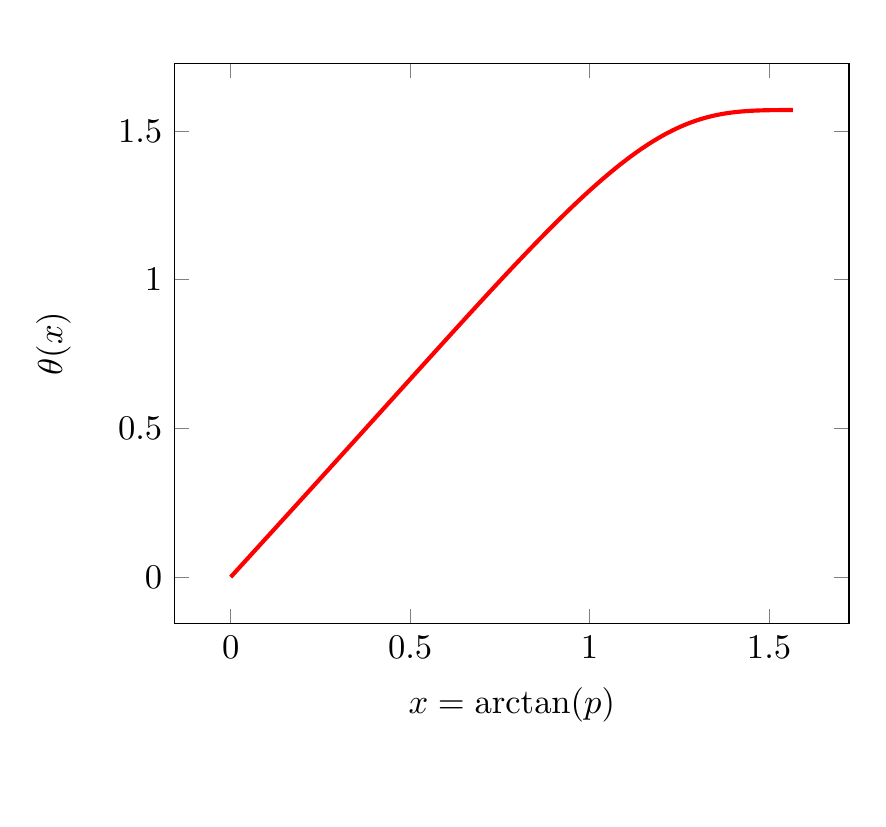}
\caption{\small 
Numerical solution of the gap equation for the Bogolyubov angle $\theta(p)$, $p$ is in units of $\sqrt{\lambda}$. 
Where x comes from the change of variable $p = \tan(x)$.}
  \label{changle}
  \end{center}
\end{figure}

Equation (\ref{ABgapeq}) can be solved at exceedingly small but finite values of $\mu_{IR}$;
then extrapolation to the limit $\mu_{IR} \rightarrow 0$ must be performed. The equation is solved recurrently
with a special care for the numerical integration in the vicinity of $ p = k$. 
The resulting chiral angle is consistent with 
previous studies \cite{Ming1,Lenz1} and is presented in Fig. \ref{changle}.

\section{\label{hsmes} The heavy-light mesons}

\subsection{Equation for the heavy-light mesons}

The Bethe--Salpeter equation for the heavy-light mesons in the laboratory frame 
follows from \cite{Bars1} in a straightforward manner, by taking the limit $m_2\to\infty$ in the coupled equations
of \cite{Bars1}, which untangles them.
The corresponding Bethe--Salpeter equation was obtained e.g. in Refs. \cite{NefKalash,msh};
an alternative derivation can be found in the text \cite{ShifmanTB}.
It has the form
%
\beqn
{\cal E} \phi (p) &=& p\,\sin \theta_p\, \,\phi(p) \nonumber\\[3mm]
&-&\lambda\,\int\, \frac{dk}{(p-k)^2}\left[ \cos\frac{\theta_p-\theta_k}{2}
\,\phi (k) \right.
\nonumber\\[3mm]
&-&
\left.\left(\cos\frac{\theta_p-\theta_k}{2}
\right)^2 \,\,\phi(p)\right] .
\label{bettereq}
\eeqn

It is not difficult to derive the boundary conditions  on $\phi(p)$ and
some properties of the wave function: 

(i) it can be taken real, nonsingular, and either symmetric or
antisymmetric under $p\to -p$,
$$
\phi (-p) =\pm \phi (p);
$$ 
and 

(ii) at large $|p|$
\beq
\phi (p ) \sim \left\{
 \begin{array}{c}
\frac{1}{|p|^3}\quad  \mbox{symmetric levels}\,,
\\[3mm]
\frac{1}{p^4}\quad \mbox{antisymmetric levels}
\end{array}
\right.
.\label{boundcond}
\eeq
This asymptotic behavior is necessary
to guarantee the cancellation of the leading (at large $p$) term on the right-hand side
of Eq.~(\ref{bettereq}).

\begin{figure}[!htpb]
  \begin{center}
    \includegraphics[width=0.4\linewidth]{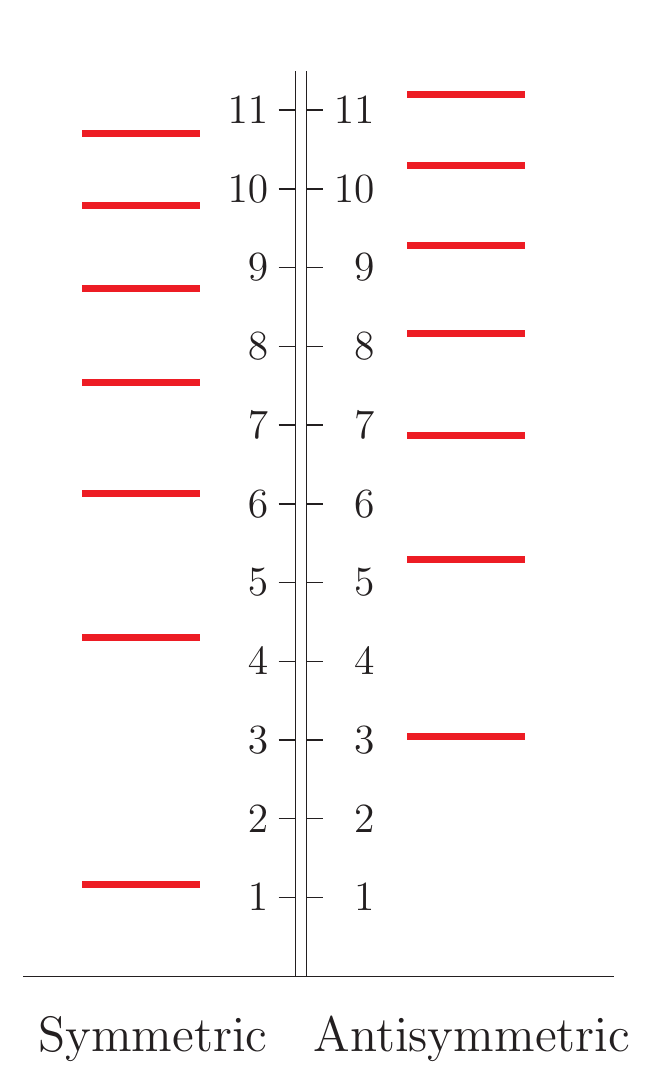}
\caption{\small 
Spectrum of the heavy-light mesons in units of $\sqrt{\lambda}$.}
  \label{spectrum}
  \end{center}
\end{figure}
\begin{figure}[!htpb]
   \includegraphics[width=\linewidth]{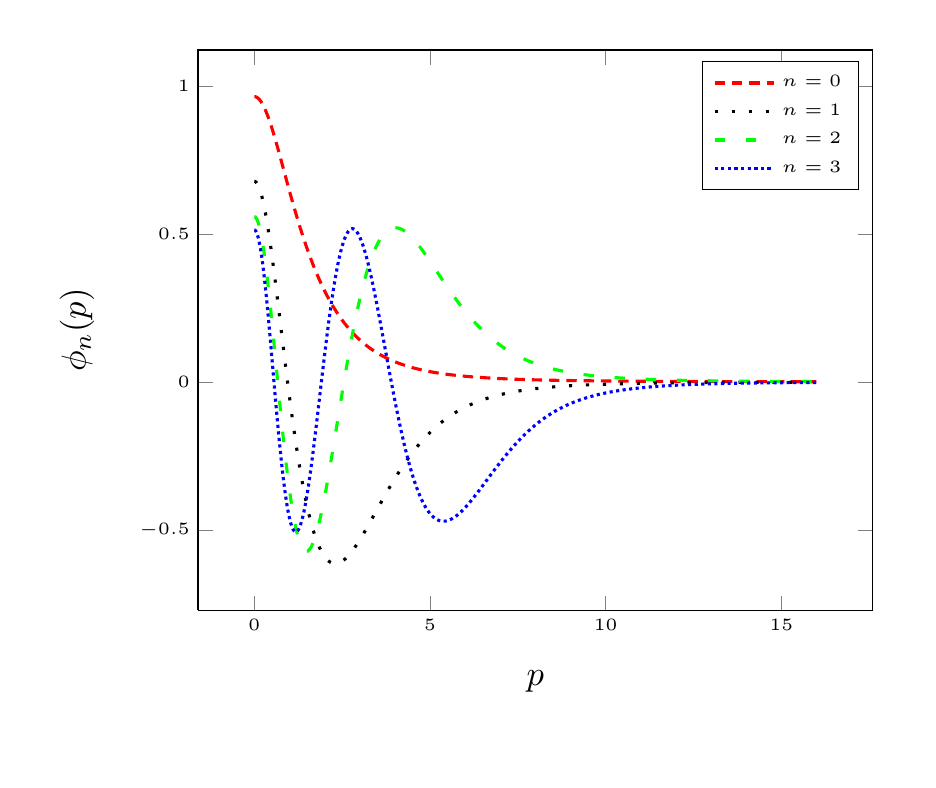}
  \caption{\small Wave functions of mesons with the negative parity (i.e. with the "symmetric" relative motion wave function).
The wave function $\phi_{n}(p)$ is in units of $\lambda^{(-1/4)}$ and momentum $p$ is in units of $\sqrt{\lambda}$.}
\label{wpion}
\end{figure}
\begin{figure}[!htpb]
  \includegraphics[width=1\linewidth]{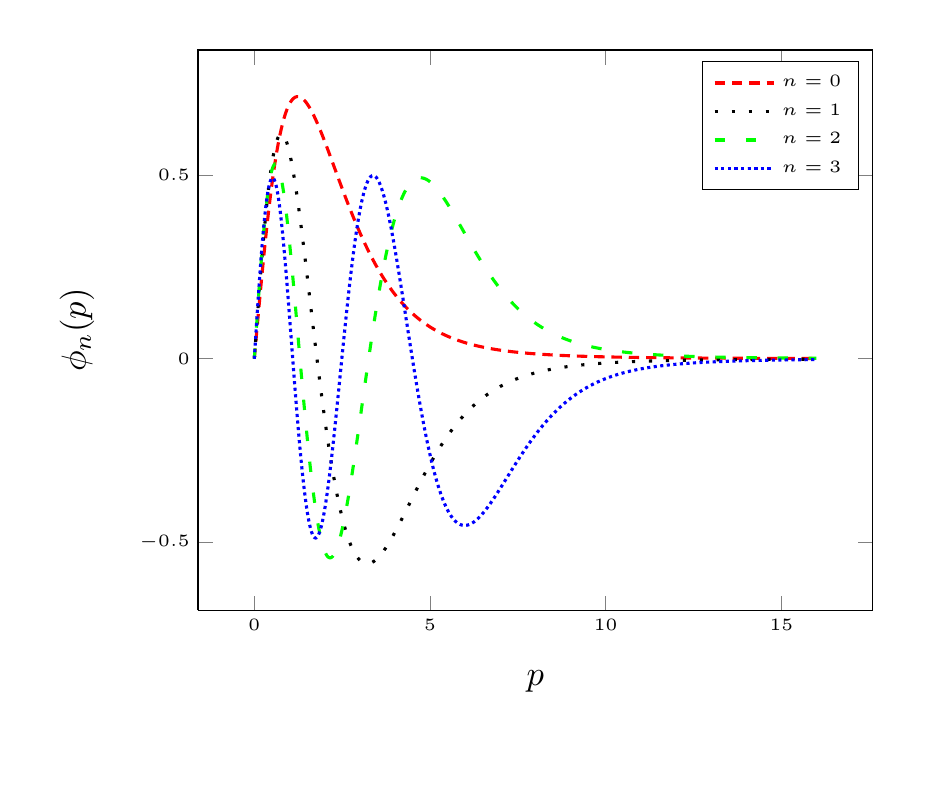}
  \caption{\small Wave functions of mesons with the positive parity (i.e. with the "antisymmetric" relative motion wave function).
The wave function $\phi_{n}(p)$ is in units of $\lambda^{(-1/4)}$ and momentum $p$ is in units of $\sqrt{\lambda}$.}
\label{wsigma}
\end{figure}

Knowing the numerical solution for the chiral angle $\theta_p$, we are able to solve equation (\ref{bettereq}). 
For the numerical solution of equation (\ref{bettereq}) it is convenient to use the regularization (\ref{scdef}).
Equation (\ref{bettereq}) then takes the form
\beqn
  {\cal E} \phi(p) &=& p \sin\theta_p \phi(p) \nonumber\\[3mm]
&-& \lambda \int \frac{dk}{(p - k)^2 + \mu_{IR}^2}
  \left[
    \cos\frac{\theta_p - \theta_k}{2}~\phi(k) 
    \right.
    \nonumber\\[3mm]
    &-&
    \left.
     \left(\cos\frac{\theta_p - \theta_k}{2}\right)^2 \phi(p)
  \right]
\, .	\label{bettereqmu}
\eeqn
Considering (\ref{bettereqmu}) at $\mu_{IR} \rightarrow 0$ one can see that all 
infrared divergences cancel each other
\begin{equation}
  {\cal E} \phi(p) = p \sin\theta_p \phi(p) - \frac{\lambda \pi}{\mu_{IR}}\phi(p) + \frac{\lambda \pi}{\mu_{IR}}\phi(p)
  + {\rm a~finite~part}\, .	\label{bsecanc}
\end{equation}
We solve Eq. (\ref{bettereqmu}) variationally by expanding the unknown wave function in the basis
\begin{equation}
  \phi(p) = \sum_{i = 1}^{N} C_i \chi_i(p)\, .	\label{chibasis}
\end{equation}
For the symmetric levels, we choose a basis in the form $$\chi_i(p) = \exp(-\alpha_i p^2)$$ 
while for antisymmetric $$\chi_i(p) = p\, \exp(-\alpha_i p^2)\,.$$
A relatively small number of gaussians
is required for a sufficiently accurate expansion.
Given the above basis, Eq. (\ref{bettereqmu}) transforms into a system of linear equations
\begin{widetext}
\begin{eqnarray}
\nonumber
  {\cal E} \sum_{i = 1}^{N} C_i \chi_i(p) &=& p\, \sin\theta_p \sum_{i = 1}^{N} C_i \chi_i(p) \\
\nonumber\\[3mm]
   &-& \lambda \int \frac{dk}{(p - k)^2 + \mu_{IR}^2}
  \left[
    \cos\frac{\theta_p - \theta_k}{2}~\sum_{i = 1}^{N} C_i \chi_i(k) 
   - 
  \left(\cos\frac{\theta_p - \theta_k}{2}\right)^2 \sum_{i = 1}^{N} C_i \chi_i(p)
  \right]\, .
\label{beqmus}
\end{eqnarray}
\end{widetext}
Multiplying (\ref{beqmus}) by $\chi_j(p)$, we obtain the generalized eigenvalue problem:
\begin{equation}
\nonumber  {\cal E} D \vec{C_{n}} = (A + B)\vec{C_{n}}\, ,
\end{equation}
where
\begin{eqnarray}
  D_{ij} &=& \int dp\, \chi_i(p) \chi_j(p)\, ,
\nonumber\\[3mm] 
  A_{ij} &=& \int dp\, p \sin\theta_p \chi_i(p) \chi_j(p)\, ,\label{geigenpr}\\[3mm] 
  B_{ij} &=& \int dp \int dk 
  \left[
    \cos\frac{\theta_p - \theta_k}{2}~\chi_i(k) \chi_j(p) 
\right.
\\
  &-&
\left.
\left(\cos\frac{\theta_p - \theta_k}{2}\right)^2 \chi_i(p) \chi_j(p)
  \right]\, .\nonumber	
\end{eqnarray}
Energy levels obtained by solving the problem (\ref{geigenpr}) 
are shown in Table \ref{tablmes} and in Fig. \ref{spectrum}, 
the corresponding wave functions are in Fig. \ref{wpion} and Fig. \ref{wsigma}.
All wave functions are normalized by condition $\int dp\, \phi^{2}(p) = 1$.

\begin{center}
\begin{table}
\caption{Energy levels of the heavy-light hadrons in units of $\sqrt{\lambda}$}
\label{tablmes}
\begin{center}
~\newline~\newline
\begin{tabular}{|c|c|c|}
 \hline
 $n$  & $P = -$ & $P = +$ \\ \hline
 $0$ & $1.161$ & $3.043$\\
 $1$ & $4.300$ & $5.286$\\
 $2$ & $6.126$ & $6.868$\\
 $3$ & $7.540$ & $8.159$\\
 $4$ & $8.734$ & $9.276$\\
 $5$ & $9.789$ & $10.27$\\ 
 $6$ & $10.74$ & $11.18$\\ \hline
\end{tabular}
\end{center}
\end{table}
\end{center}

\newpage
\subsection{The heavy-light mesons on the light cone}

Now we deal with the 't Hooft-like equation (\ref{HQLtHooft}).
In order to solve it numerically we split the integral into two parts
\begin{eqnarray}
  2 {\cal E}_{m}\varphi_m(\xi) &=& \sqrt{2 \lambda}\, \xi\, \varphi_m(\xi) \nonumber\\[3mm]
  &-&\sqrt{2 \lambda}\, \lim_{\epsilon \rightarrow 0} 
  \left( 
    \int_{0}^{\xi - \epsilon} \frac{\varphi_m(\tilde{\xi}) - \varphi_m(\xi)}{(\tilde{\xi} - \xi)^2} d\tilde{\xi} \right. \nonumber\\[3mm]
    &+&
    \left.\int_{\xi + \epsilon}^{\infty} \frac{\varphi_m(\tilde{\xi}) - \varphi_m(\xi)}{(\tilde{\xi} - \xi)^2} d\tilde{\xi} 
  \right)
\, .	\label{Zhitneps}
\end{eqnarray}

Alternatively the 't Hooft-like equation can be solved with definition (\ref{scdef}). 
Then it takes form:
\begin{equation}
  2 {\cal E}_{m}\varphi_m(\xi) = \sqrt{2 \lambda}\, \xi\, \varphi_m(\xi) - \sqrt{2 \lambda} \int_{0}^{\infty}
   \frac{\varphi_m(\tilde{\xi}) - \varphi_m(\xi)}{(\tilde{\xi} - \xi)^2 + \mu_{IR}^2} d\tilde{\xi}\, ,	\label{Zhitnmu}
\end{equation}
where $\mu_{IR} \rightarrow 0$ is assumed.

\begin{figure}[!htpb]
  \begin{center}
    \includegraphics[width=1\linewidth]{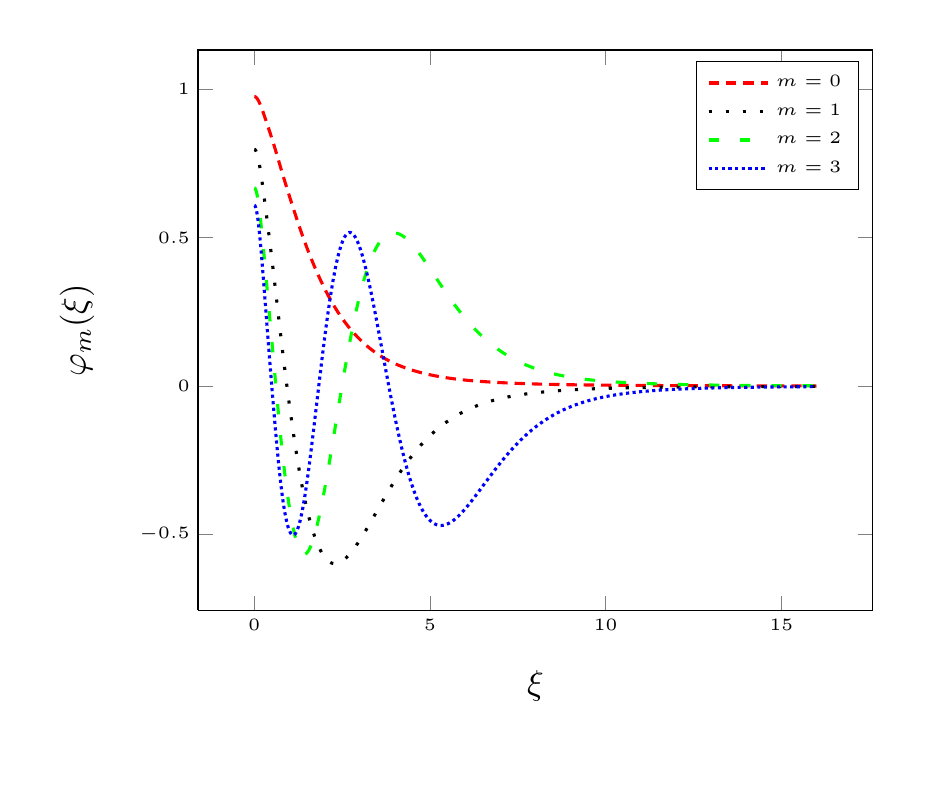}
\caption{\small 
Wave functions of mesons obtained from the 't Hooft-like equation.
Even $m$ represent the negative parity mesons and odd $m$ correspond to the positive parity mesons.
Both the wave functions $\varphi_{m}(\xi)$ and the variable $\xi$ are dimensionless.}
  \label{ZhitnPlot}
  \end{center}
\end{figure}

Both equations (\ref{Zhitneps}) and (\ref{Zhitnmu}) were solved numerically in much the same way
as Eq.~(\ref{bettereqmu}).
The results in both cases (\ref{Zhitneps}) and (\ref{Zhitnmu}) coincide. The spectrum is identical 
to that following from the laboratory-frame equation (\ref{bettereqmu}), see Fig.~\ref{spectrum}.
The light-cone wave functions are normalized by the condition $\int d\xi\, \varphi_{m}^{2}(\xi) = 1$ and presented in Fig. \ref{ZhitnPlot}.

\subsection{Equation (\ref{bettereq}) with the unphysical chiral angle vs. the
't Hooft-like equation}
\label{unphis}

People are used to the fact that the chiral condensate cannot be directly captured if one works on the light cone.
At the same time, the chiral symmetry breaking is seen indirectly, through the absence of the parity degeneracy
in the spectrum of physical mesons. The situation in our laboratory-frame construction is totally different.
The nonsingular solution for $\theta_p$, see Section \ref{NGapSolSec}, immediately produces 
$\langle\bar\psi\psi\rangle\neq 0$, see Eq. (\ref{psibarpsii}). As a result, naturally, all $P$-odd states split from $P$-even.

The singular solution (\ref{unsst}) would lead to $\langle\bar\psi\psi\rangle =0\, .$ If, using (\ref{unsst}), we could
obtain a consistent laboratory-frame Bethe--Salpeter equation, with a proper Foldy--Wouthuysen
transformation, it should have produced  a parity degenerate meson spectrum,
in full accord with general theorems. However, (\ref{unsst}) implies (\ref{unsse}), which obviously
precludes the use of (\ref{unsst}) in the Bethe--Salpeter equation
because of negativity of the solution (\ref{unsse}) at small $|p|$.

Physically it means that the chiral symmetry
is {\em a priori}  broken in the 't Hooft model. Trying to restore it by brute force insisting on the
chirally symmetric vacuum, we see that the bound state equation for hadrons in the rest frame
is not defined, and no consistent solutions for hadronic spectrum exists.

Nevertheless, let us perform this incorrect and illegitimate operation, and see what happens. Below
we examine a strange construct, namely, Eq.~(\ref{bettereq}) with the {\em singular} (unphysical) chiral angle, i.e.
we replace $\theta_{p,k}$  in Eq.~(\ref{bettereq}) by (\ref{unsst}). This is no longer a legitimate
laboratory-frame Bethe--Salpeter equation. But it has a miraculous feature.

For positive values of $p$ we get
\beq
{\cal E} \phi (p) =  p\, \phi(p)
-\lambda\,\int_0^\infty\, \frac{dk}{(p-k)^2}\left[ 
\,\phi (k) -
 \,\,\phi(p)\right] .
\label{bettequ}
\eeq
Next, we introduce dimensionless variables (marked by tildes)
\beq
p =\sqrt{\lambda}\, \tilde{p}\,,\qquad k =\sqrt{\lambda} \,\tilde{k}\,.
\label{dv}
\eeq
The wave functions are to be understood now as functions depending on $\tilde{p},\,\,\tilde{k}$ rather than 
$p,\,\,k $, although we will keep using the same notation $\phi$.
Then, in terms of these dimensionless variables, Eq.~(\ref{bettequ}) takes the form
\beq
{\cal E} \phi (p) =  \sqrt{\lambda} \,\tilde{p}\,  \phi(\tilde{p})
-\sqrt\lambda\,\int_0^\infty\, \frac{d\tilde{k}}{(\tilde{p}-\tilde{k})^2}\left[ 
\,\phi (\tilde{k}) -
 \,\,\phi(\tilde{p})\right] .
\label{bettequp}
\eeq
Compare it with Eq. (\ref{Zhitneps}) or (\ref{Zhitnmu}). We observe, with surprise, that Eq. (\ref{bettequp}) is
identical to (\ref{Zhitneps}), up to a renaming of the integration variables and rescaling
\beq
{\cal E} \to \sqrt{2} {\cal E}_m\, .
\eeq
Thus, the laboratory frame Bethe--Salpeter equation with the {\em wrong} chiral angle
and the boundary conditions inappropriate for the laboratory frame equation
\footnote
{
The laboratory frame Bethe--Salpeter equation requires vanishing of the odd wave functions at $p=0$, while the wave functions of the 
't Hooft-like equation do not vanish at $\xi=0$.
}
reproduces the spectrum of the ({\em correct}) 't Hooft-like light-cone equation
up to an overall energy scale which is off by a factor of $1/\sqrt{2}$. In particular, the ratios of the energy levels
following from (\ref{bettequ}) are correct. The physical reason for this coincidence remains puzzling. 


~\newline
\section{\label{concl} Conclusions}

We studied the heavy-light mesons in $(1+1)$-dimensional QCD in the 't Hooft limit, with the emphasis on the 
impact of the chiral symmetry breaking both on the spectrum and wave functions. To this end we compared 
two alternative quantization schemes:  laboratory frame Bethe--Salpeter equation with a nontrivial chiral angle
and  the light-cone 't Hooft-like equation which has no direct information on the chiral condensate in the vacuum.
Two distinct limiting procedures leading to these two respective equations are not a priori interchangeable.
 
First, we   solved the system in the laboratory frame using the Coulomb (axial) gauge.
The solution proceeds via two steps. One begins from the solution of  the gap equation and obtains a single-quark 
Green's function as well as the quark condensate in the vacuum. Chiral symmetry is manifestly dynamically broken
in the vacuum. Then one solves the Bethe--Salpeter equation determining the odd and even wave functions and the spectrum.
Chiral symmetry is broken in the spectrum too. The spectral results  are independent
on the gauge choice and on an infrared regularization scheme.

Second, we  solved the same system on the light cone. In this case there is no analog of the gap equation,
and vacuum is trivial.  Nevertheless, the chiral symmetry is   broken in the observable  
spectrum. 
Needless to say, all wave functions are totally different (they depend on variable which have very different meanings in these two schemes).
While dynamical equations on the light cone and in the laboratory frame (with the Coulomb gauge)
look very different, the results for the spectra are the same. We demonstrated this numerically;
 the question of explicitly  finding an appropriate unitary transformation between both schemes remains open.

A curious fact was observed {\em en route}.
The
laboratory frame equation for $\phi (p)$   becomes identical to the 't
Hooft-like equation for 
$\varphi (\xi )$ (see Eq. (\ref{HQLtHooft})) upon substitution into the laboratory-frame equation
 a singular (nonphysical) solution for the chiral angle with simultaneous rescaling of the overall
energy scale. 

{\bf Acknowledgements}
L.Ya.G. and V.K.S. acknowledge support from the Austrian Science
Fund (FWF) through the grant P21970-N16. M.S. is grateful to 
Frieder Lenz and Michael Thies for useful discussions. The work of M.S.
is supported in part by  DOE grant DE-FG02-94ER40823.



\begin{thebibliography}{99}

\bibitem{tHoof1}
G. 't Hooft, {\it Nucl. Phys. } {\bf B75} (1974) 461.

\bibitem{Call1}
C. G. Callan, N. Coote, and D.J. Gross,
{\it Phys. Rev.} {\bf D13} (1976) 1649.

\bibitem{Einh1}
M. B. Einhorn,  {\it Phys. Rev.} {\bf D14} (1976) 3451.
  
\bibitem{Hornbostel:1988ne}
K.~Hornbostel, {\em The Application Of Light-Cone Quantization To Quantum Chromodynamics In (1+1)-dimensions},
 SLAC Ph.D. Thesis, 1988.

\bibitem{Zhit1}
A. Zhitnitsky, {\it Phys. Lett.} {\bf B 165} (1985) 405;
{\it Sov. J. Nucl. Phys. } {\bf 43} (1986) 999; {\bf 44} (1986) 139.
  
\bibitem{Ming1}
Ming Li, {\it Phys. Rev. } {\bf D34} (1986) 3888;
  \\
Ming Li, L. Wilets,  and M. C. Birse,
{\it J. Phys.} {\bf G13} (1987) 915.
  
\bibitem{Lenz1}
F. Lenz, M. Thies, S. Levit, and K. Yazaki,
{\it Ann. Phys. (N.Y.)} {\bf 208} (1991) 1.

\bibitem{Burk1}
M. Burkardt, in Proc. of the Workshop on
Quantum Infrared Physics, Paris, France, 6-10 June 1994,
{\it  Quantum Infrared Physics},  Eds.   H.M.
Fried and B. Muller (World Scientific, 1995)
[hep-ph/9409333]. 

\bibitem{Bars1}
I. Bars and M.B. Green,  {\it Phys. Rev. } {\bf D17} (1978) 537.

\bibitem{Grin1}
B. Grinstein and P. Mende,
{\it Nucl. Phys.} {\bf B425} (1994) 451,
  [hep-ph/9401303].

\bibitem{Bur}
 M.~Burkardt and E.~S.~Swanson,
  Phys.\ Rev.\ D {\bf 46}, 5083 (1992).

\bibitem{Zhit2}
A. Zhitnitsky, {\it Phys. Rev.} {\bf D53} (1996) 5821, [hep-ph/9510366].

\bibitem{Bjor1}
J. D. Bjorken, {\it Phys. Rev.} {\bf D17} (1978) 171.

\bibitem{NefKalash}
Yu. S. Kalashnikova, A. V. Nefediev,   
 {\it Phys. Atom. Nucl.} {\bf 62} (1999) 323,
  [hep-ph/9711347];
Yu. S. Kalashnikova, A. V. Nefediev, {\it   Phys.\ Usp.} {\bf 45}, 4 (2002) 347,
  [hep-ph/0111225].
  
  \bibitem{msh}
  M. Shifman, {\em Light-Cone, Chiral Condensates,  Heavy Quarks and the 't Hooft
Model}, 
Talk at the Sixth International  Workshop on Light-Front
Quantization and  Non-Perturbative QCD,   June 3 -- 14, 1996,
International Institute  of  Theoretical and  Applied  Physics,
Iowa  State  University, Ames, Iowa.

\bibitem{ShifmanTB}
M. Shifman,  {\em Advanced Topics in Quantum Field Theory}, (Cambridge University Press, 2012), Section
41.
\end{thebibliography}
\end{document}